\documentclass{jps-cp}

\usepackage{bm}
\usepackage{color}
\usepackage{comment}
\usepackage{amsmath}
\usepackage{rsfs}

\usepackage{txfonts} %Please comment out this line unless the txfonts package is availabe in your LaTeX system.
\usepackage{tikz}

\title{
Quark confinement due to unified magnetic monopoles and vortices reduced from symmetric instantons with holography
}

\author{Kei-Ichi \textsc{Kondo}$^{1,2}$}

\inst{$^{1}$%Department of Physics, 
Graduate School of Science, Chiba University, 1-33 Yayoi-cho, Chiba, Chiba 263-8522, Japan \\
$^{2}$Research and Education Center for Natural Sciences, Keio University, 
Kanagawa 223-8521, Japan}

\email{kondok@faculty.chiba-u.jp}

\recdate{November 30, 2025}

\abst{
We develop a geometric framework to analyze quark confinement in four-dimensional Euclidean
$SU(2)$ Yang--Mills theory in terms of finite-action topological defects.
Starting from self-dual Yang--Mills configurations, we restrict to \emph{symmetric instantons} with
spatial rotation symmetry so that dimensional reduction preserves conformal equivalence.
This requirement maps $\mathbb{R}^4$ to curved backgrounds with compact directions and, in particular,
identifies the reduced configurations with (i) hyperbolic magnetic monopoles of Atiyah type on
$H^3\simeq \mathrm{AdS}_3$ (from an $SO(2)\simeq S^1$ symmetry) and (ii) hyperbolic vortices of
Witten--Manton type on $H^2\simeq \mathrm{AdS}_2$ (from an $SO(3)\simeq SU(2)$ symmetry).
We provide an explicit field map relating the monopole and vortex variables, enabling a unified
treatment of these defects within the original four-dimensional setting.
Moreover, the hyperbolic monopole on $H^3$ is completely determined by its holographic data on the
conformal boundary $S^2_\infty$, which reduces a non-Abelian Wilson loop placed on $\partial H^3$
to an Abelian loop determined by the vortex $U(1)$ field (Abelian dominance and monopole dominance),
without further dynamical assumptions beyond the symmetry reduction.
In the semiclassical dilute-gas regime of these finite-action defects, the framework yields the Wilson
area law, thereby providing analytic support for the dual-superconductor picture of confinement.
}

\kword{quark confinement, instanton, magnetic monopole, vortex, holography %\ldots
}

\begin{document}
\maketitle

\section{Introduction}
%\noindent
%\textcolor{green}{\Large\bf $\S$~~Introduction
%}

%\noindent
%$\bullet$ 
\textbf{Quark confinement} means that quarks as the most fundamental building blocks of the matter have never been observed in the isolated form and  must be confined in hadrons.
%
%\noindent
%$\bullet$ 
This is caused by \textbf{strong interactions} mediated by gluons which are described by the \textbf{Yang-Mills theory}, i.e., the non-Abelian gauge theory.
%
%\noindent
%$\bullet$ 
In this talk we consider \textbf{quark confinement} in the 4-dim. ($D=4$) quantum Yang-Mills theory according to the \textbf{Wilson criterion} (with no dynamical quarks): 
%
%\noindent
\\
\quad
\textbf{area law} of the Wilson loop average 
$\Leftrightarrow$  \textbf{linear potential} for static $q\bar{q}$ %quark--anti-quark 
potential.

%\noindent
%$\bullet$ 
Quark confinement in this sense can be understood based on the \textbf{dual superconductor picture} proposed by Nambu, `t Hooft, Mandelstam, Polyakov in the mid-1970s. 
For this purpose, we need \textbf{magnetic monopoles} and/or  \textbf{vortices}.
For a review, see e.g., Kondo et al.\cite{PR}.
%
%\noindent%
%$\bullet$ 
Nevertheless, topological solitons in Yang-Mills theory are only  \textbf{instantons} in the $D=4$ Euclidean space $\mathbb{R}^4$.
%
%\noindent
It is a big question how to derive such lower-dim. topological objects from the $D=4$ Yang-Mills theory.

%\begin{comment}
%\newpage
%\noindent
%\textcolor{green}{\Large\bf $\S$~~Introduction and our strategy
%}
%\setcounter{equation}{0}
%\normalsize 

%\noindent
%$\bullet$ Superconductivity
%\\
%Magnetic charge $\to$ squeezed flux of magnetic field (Abrikosov vortex)  $\leftarrow$ Meissner effect [Field affected: gauge field]
%\\
%Vacuum condensation of electric object (Cooper pair 2e) [Field that creates condensation: scalar field]

%\noindent
%$\bullet$ \textcolor{red}{Dual superconductivity} if electro-magnetic duality holds:
%\\
%Color electric charge $\to$ squeezed flux of color electric field (\textbf{vortex}?)  $\leftarrow$ dual Meissner effect?
%Vacuum condensation of magnetic object (\textbf{magnetic monopole}?)?

%For this description, not only gauge field but also scalar field is required. (Remember the Ginzburg-Landau theory), e.g., 
%Nielsen-Olesen vortex, `tHooft-Polyakov magnetic monopole.
%However, Yang-Mills theory does not have scalar field, only the gauge field exists.
%\end{comment}

%\begin{comment}
%\newpage
%\noindent
%$\odot$ 
The topological solitons in the Yang-Mills theory are only \textbf{instantons} in 4-dim. Euclidean spacetime $\mathbb{R}^4$. [Coleman-Deser-Pagels theorem]
It is known that various low-dimensional integrable equations can be obtained from the self-dual Yang-Mills equations in 4-dim.  space $\mathbb{R}^4$ by dimensional reduction.
%KdV equation, KP equation, sine-Gordon equation, NLS equation, Liouville equation, etc. . .
%
%\noindent
%$\odot$ 

The $D=4$ Yang-Mills theory  has \textbf{conformal symmetry}. %(invariant under \textbf{conformal transformation}). 
The self-dual Yang-Mills equation on $\mathbb{R}^4$ has also the conformal symmetry, whose solutions (instantons) give solutions of the Yang-Mills field equation with a finite Euclidean action. 
%
%\noindent
Therefore, we consider the Yang-Mills theory on the 4-dim. Euclidean spacetime $\mathbb{R}^4(x^1,x^2,x^3,x^4)$. 
%\\
%\\
%For example, if we perform dimensional reduction to eliminate Euclidean time $t$,
%\\
%$\mathscr{A}_\mu(x^1,x^2,x^3,{t}=t) \to \mathscr{A}_\mu(x^1,x^2,x^3)=(\mathscr{A}_k(x^1,x^2,x^3),\Phi(x^1,x^2,x^3))$,
%\\
%The 4-dimensional self-dual equation reduces to a 3-dimensional magnetic monopole equation. The solution is a 3-dimensional magnetic monopole.
%\\
%\noindent
%$\odot$ 
%However, if we substitute the 3-dimensional magnetic monopole solution into the action of the 4-dimensional Yang-Mills theory, the action diverges from the integration with respect to time because this solution does not depend on time. Therefore, the 3-dim.  magnetic monopole configuration obtained in this way does not contribute to the path (functional) integral of the 4-dim.  Yang-Mills theory, and the magnetic monopole does not act as a confinement mechanism.
%\\
%\noindent
%$\odot$ 
%The same can be said for other low-dim. topological solitons, such as vortices.
%One way to avoid this is to appropriately compactify space-time ($\mathbb{R}^4 \to \mathbb{R}^3 \times S^1, \mathbb{R}^2 \times S^1 \times S^1=\mathbb{R}^2 \times T^2, \mathbb{R}^2 \times S^2,...$), but %are there 
%any criteria for choosing it?
%\newpage
%
%\noindent
In this talk we show that \textbf{$D=3$ magnetic monopoles and $D=2$ (center) vortices responsible for quark confinement are constructed from symmetric instantons in the $D=4$ Euclidean Yang-Mills theory} in a way consistent with \textbf{holography principle}. 
%\textcolor{blue}{
%In this talk we show that both objects arise from a single class of symmetric instantons, leading to a unified and holographically controlled confinement mechanism.
%}
%
%\noindent
This result is obtained \cite{Kondo25} based on the guiding principles: 
\\
\noindent
%$\bullet$ \textcolor{red}{conformal equivalence}:
$\bullet$ {conformal equivalence}: conformal symmetry, 
\\
\noindent
%$\bullet$ \textcolor{red}{symmetric instanton gauge field}:
$\bullet$ {symmetric instanton gauge field}: spatial symmetry $SO(2)$, $SO(3)$,
\\
\noindent
%$\bullet$ \textcolor{red}{dimensional reductions}:
$\bullet$ {dimensional reductions}: self-dual equation (electric-magnetic dual symmetry).

\section{Translation symmetry and dimensional reduction}
%\noindent
%\textcolor{green}{\Large\bf $\S$~Translation symmetry and dimensional reduction
%}
%\setcounter{equation}{0}
\normalsize 
%\small
%\\
%$\odot$ 
We consider $SU(2)$ Yang-Mills theory on the $D=4$ Euclidean space $\mathbb{R}^4(x^1,x^2,x^3,{t})$.
The Euclidean time $x^4$ is written as $t$ in what follows. 
%\begin{align}
%&\mathscr{L} = \frac12 {\rm tr}(\mathscr{F}_{\mu\nu}(x) \mathscr{F}_{\mu\nu}(x)) , \ x = (x^1,x^2,x^3,{t})=(\bm{x},t) \in \mathbb{R}^4 , 
%%\nonumber
%\\
%&\mathscr{F}_{\mu\nu}(x) := \partial_\mu \mathscr{A}_\nu(x) - \partial_\nu \mathscr{A}_\mu(x)  - ig [\mathscr{A}_\mu(x), \mathscr{A}_\nu(x)] , \ \mathscr{A}_\mu(x) := \mathscr{A}_\mu^A(x) \frac{\sigma_A}{2} .
%%\nonumber
%\end{align}
For the Yang-Mills field $\mathscr{A}_\mu(x) := \mathscr{A}_\mu^A(x) \frac{\sigma_A}{2} $ with the Pauli matrices $\sigma_A (A=1,2,3)$, the action is given by
%we consider the $D=4$ Euclidean $SU(2)$ Yang-Mills theory with the action given by 
\begin{align}
S_{\rm{E}}^{\rm{YM}} =& \int d^4 x~\mathscr{L}_{\rm{E}}^{\rm{YM}} 
= \int d^4 x \frac12 {\rm tr}(\mathscr{F}_{\mu\nu}(x) \mathscr{F}_{\mu\nu}(x))
= \int d^4 x~\frac{1}{4}\mathscr{F}_{\mu\nu}^A(x) \mathscr{F}_{\mu\nu}^A(x) , %\geq 0 ,
%\end{align}
%\begin{align}
%&\mathscr{L} = \frac12 {\rm tr}(\mathscr{F}_{\mu\nu}(x) \mathscr{F}_{\mu\nu}(x)) , \ x = (x^1,x^2,x^3,{t})=(\bm{x},t) \in \mathbb{R}^4 , 
\nonumber\\
%\\
 \mathscr{F}_{\mu\nu}(x) :=& \partial_\mu \mathscr{A}_\nu(x) - \partial_\nu \mathscr{A}_\mu(x)  - ig [\mathscr{A}_\mu(x), \mathscr{A}_\nu(x)] 
 = \mathscr{F}_{\mu\nu}^A(x) \frac{\sigma_A}{2} \in su(2). 
%%\nonumber\\
%\mathscr{A}_\mu(x) :=& \mathscr{A}_\mu^A(x) \frac{\sigma_A}{2} \ (A=1,2,3) .
%%\nonumber
\end{align}
%with the flat metric 
%$%\begin{equation}
% (ds)^2(\mathbb{R}^4) =  (dx^1)^2 + (dx^2)^2 + (dx^3)^2 + (d{t})^2 .
%$%\end{equation}
In what follows use $g$ for the gauge coupling constant $g_{{}_{\rm YM}}$  to simplify the notation. 
The the metric is given by
\begin{equation}
 (ds)^2(\mathbb{R}^4) =  [(dx^1)^2 + (dx^2)^2 + (dx^3)^2] + (d{t})^2 
\Longrightarrow  \mathbb{R}^4 = \mathbb{R}^3 \times \mathbb{R}^1
.
%\begin{equation}
%\begin{array}{cccccccc}
%\mathbb{R}^4 = \mathbb{R}^3 \times \mathbb{R}^1\rightarrow&\mathbb{R}^4  &\backslash& \mathbb{R}^2 &\simeq&  \mathbb{H}^3& \times &S^1\\
%&\rotatebox{90}{$\in$}&&\rotatebox{90}{$\in$}&&\rotatebox{90}{$\in$}&&\rotatebox{90}{$\in$}\\
%&(x^1,x^2,x^3,{t})&&(x^3,{t})&&(\rho,x^3,{t})&&\varphi
%%&(x,y,z,t)&&(z,t)&&(\rho,z,t)&&\varphi
%\end{array}
%\end{equation}
%\nonumber
\end{equation}
The self-dual Yang-Mills equation is given by
\begin{align}
 *\mathscr{F}_{\mu\nu}(\bm{x},t) := \frac{1}{2} \epsilon_{\alpha\beta\mu\nu} \mathscr{F}_{\alpha\beta}(\bm{x},t) 
= \mathscr{F}_{\mu\nu}(\bm{x},t) 
\ [x = (x^1,x^2,x^3,{t})=(\bm{x},t) \in \mathbb{R}^4 ]. %%\nonumber\\
%\Leftrightarrow & 
%\pm \frac{1}{2} \epsilon_{\rho\sigma\mu\nu} \mathscr{F}_{\rho\sigma}(x^1,x^2,x^3,{t}) =\mathscr{F}_{\mu\nu}(x^1,x^2,x^3,{t}) .
%%\nonumber
\label{SD-YMeq}
\end{align}

%$\odot$ 
\noindent 
(0) 
First, we consider a solution for the gauge field that has the %translational invariance. For example, a solution that has 
\textbf{translation symmetry in the time $\textcolor{red}{t={x^4}}$}, which is equivalent to the \underline{$t$-independence}: %$(\bm{x},\textcolor{red}{t}) \to (\bm{x})$. 
%The dimensional reduction occurs: %considering a solution that does not depend on the Euclidean time $t={t}$:
%(Recall cyclic coordinates in analytical mechanics.)
\begin{align}
 (\mathscr{A}_{1}(\bm{x},t), \mathscr{A}_{2}(\bm{x},t), \mathscr{A}_{3}(\bm{x},t), \mathscr{A}_{t}(\bm{x},t)) 
 \to 
 (\mathscr{A}_{1}(\bm{x}), \mathscr{A}_{2}(\bm{x}), \mathscr{A}_{3}(\bm{x}), \Phi(\bm{x})) .
%%\nonumber
\end{align}
The time-independent solution of the self-dual equation (\ref{SD-YMeq}) %with translational invariance in the time ${t}$ direction 
reduces to the solution of \textbf{Bogomolny equation}\index{Bogomolny equation} on $\mathbb{R}^3$:
\begin{align}
% & \mathscr{F}(x^1,x^2,x^3) = *\mathscr{D}_\ell \Phi(x^1,x^2,x^3) %\nonumber\\
% \Leftrightarrow \quad & 
 (*\mathscr{F})_{\ell t}(\bm{x}) = \mathscr{D}_\ell \Phi(\bm{x}), \ \ell=1,2,3,  \quad \bm{x} := (x^1,x^2,x^3) \in \mathbb{R}^3 .
%\nonumber
\end{align}
In fact, the self-dual equation for $\mu, \nu=\ell , t$ reads for $\Phi(\bm{x}) := \mathscr{A}_{t}(\bm{x})$
%\begin{align}
% \pm \frac{1}{2} \epsilon_{jk\ell 4} \mathscr{F}_{jk}(\bm{x}) = \mathscr{F}_{\ell 4}(\bm{x}) ,
%\end{align}
%the right-hand side reads 
\begin{align}
\pm \frac{1}{2} \epsilon_{jk\ell t} \mathscr{F}_{jk}(\bm{x}) 
=\mathscr{F}_{\ell t}(\bm{x}) %%\nonumber\\
 =&  \partial_\ell \mathscr{A}_t(\bm{x}) - %\textcolor{red}
 {\partial_{t}} \mathscr{A}_\ell (\bm{x}) %%\nonumber\\& 
 -ig[ \mathscr{A}_\ell (\bm{x}), \mathscr{A}_t(\bm{x})] 
 \ (%\textcolor{red}
 {\partial_{t}} \mathscr{A}_\ell (x^1,x^2,x^3)=0)
\nonumber\\
 =& \partial_\ell \mathscr{A}_t(\bm{x}) -ig[ \mathscr{A}_\ell (\bm{x}), \mathscr{A}_t(\bm{x})]  %%\nonumber\\
 = \mathscr{D}_\ell \Phi(\bm{x}) .%, \ \Phi(\bm{x}) := \mathscr{A}_4(\bm{x}) .
%\nonumber
\end{align}
The solution of the Bogomolny equation is called the \textbf{Prasad-Sommerfield (PS) magnetic monopole}.
However, this solution leads to a divergent 4-dim.  action:
\begin{equation}
S=\int_{-\infty}^{\infty}d{t} \left[ \int dx^1 dx^2 dx^3  \mathscr{L}(x^1,x^2,x^3)\right] =\infty \Longrightarrow \exp( -S/\hbar) =0 ,
%\nonumber
\end{equation} 
even if $\int dx^1 dx^2 dx^3  \mathscr{L}(x^1,x^2,x^3)<\infty$ 
because of the $t$-independence.

Therefore,  the PS magnetic monopole does not contribute to the path integral. %because  
%\begin{equation}
%\exp( -S/\hbar) =0 .
%\end{equation} 
Thus, the PS magnetic monopole is not responsible  
%does not play any role in the mechanism 
for quark confinement. 
How to avoid this difficulty?

%\newpage
\section{Conformal equivalence (I)}
\begin{figure}[htb]
\begin{center}
\includegraphics[scale=0.40]{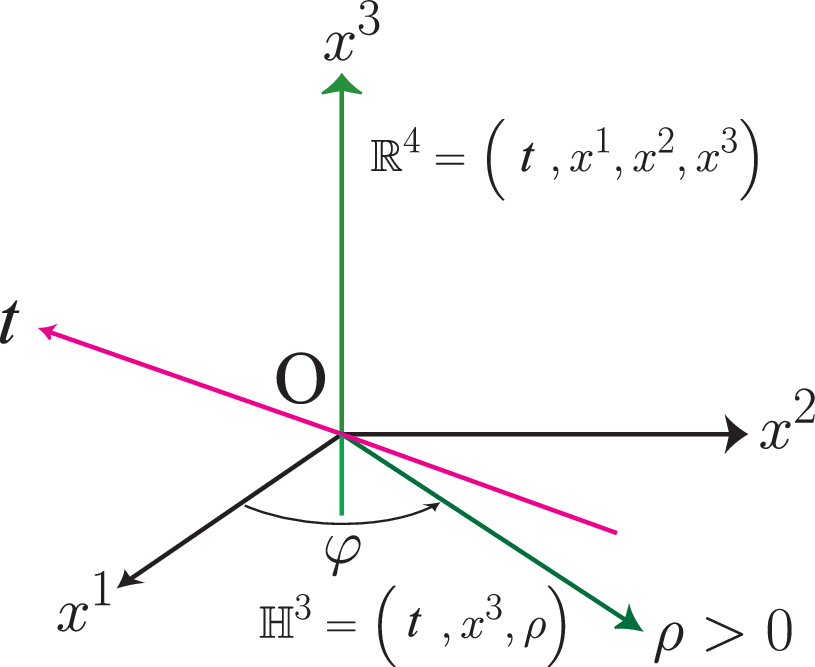}
\quad
\includegraphics[scale=0.40]{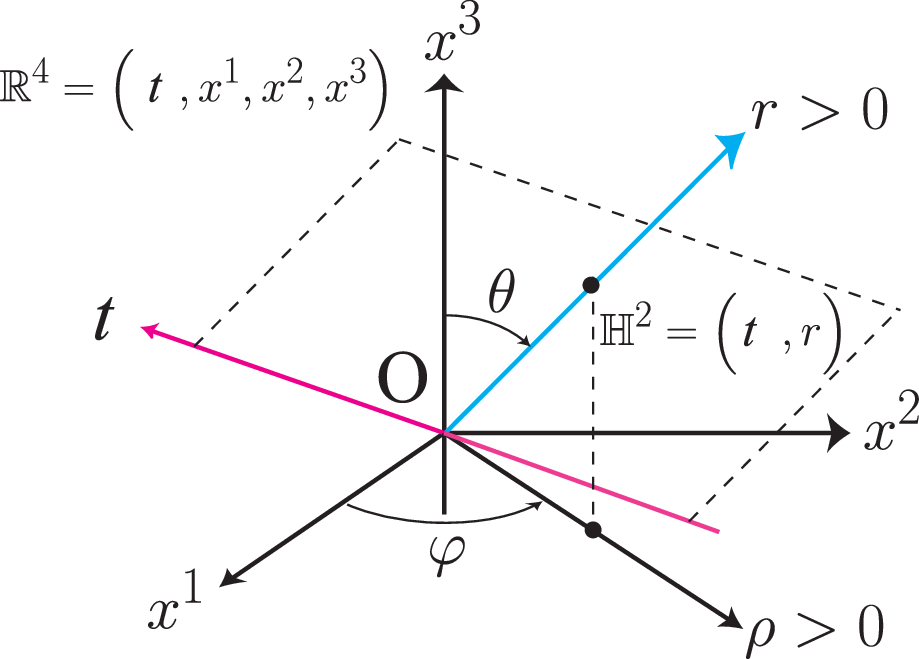}%\includegraphics[scale=0.45]{fig-ep276/hyperbolic_monopole_vortex_t.eps}
\end{center}
\caption{
Conformal equivalence and symmetry reduction: 
(Left) 4-dim. Euclidean space $\mathbb{R}^4({t},x^1,x^2,x^3)$ versus 3-dim. hyperbolic space $\mathbb{H}^3({t},x^3,\rho)$.
\\
(Right) 4-dim. Euclidean space $\mathbb{R}^4({t},x^1,x^2,x^3)$ versus 2-dim. hyperbolic space $\mathbb{H}^2({t},r)$.
%$D=3$ Euclidean space $\mathbb{R}^4({t},x^1,x^2,x^3)$ versus $D=3$  hyperbolic space $\mathbb{H}^3({t},x^3,\rho)$ with $\rho:=\sqrt{(x^1)^2+(x^2)^2} > 0$and $D=2$ hyperbolic plane $\mathbb{H}^2({t},r)$ with $r:=\sqrt{(x^1)^2+(x^2)^2+(x^3)^2}> 0$.
}
\label{conformal_equiv_all}
\end{figure}

%\noindent
%\textcolor{green}{\Large\bf $\S$~~Conformal equivalence (I)
%}
%\setcounter{equation}{0}
\normalsize 
%\small
%\\
\noindent
(I)
Next, we consider solutions with \textbf{spatial rotation symmetry}: $SO(2) \simeq S^1$.
%We start from the flat 4-dimensional Euclidean space 
\\
In %4-dim. Euclidean space 
$\mathbb{R}^4$ with the metric $(ds)^2(\mathbb{R}^4) 
= (dx^1)^2 + (dx^2)^2 + (dx^3)^2 + (d{t})^2$,  % with the Cartesian coordinates $(x^1,x^2,x^3,{t})$.
we introduce the coordinates $(\rho,\varphi)$ in the 2-dim. space $(x^1,x^2)$ to rewrite the metric:%  in the cylindrical coordinates $(\rho,\varphi,x^3,{t})$:
\begin{align}
(ds)^2(\mathbb{R}^4) 
%= (dx^1)^2 + (dx^2)^2 + (dx^3)^2 + (d{t})^2 %%\nonumber\\
= (d\rho)^2 + \rho^2(d\varphi)^2 + (dx^3)^2 + (d{t})^2 ,%\ (\rho := \sqrt{x^2 + y^2})
%%\nonumber\\
%=& \rho^2\left[\frac{(dx^3)^2 + (d{t})^2 + (d\rho)^2}{\rho^2} + (d\varphi)^2\right].
%\nonumber
%\label{R4toH3}
\end{align}
%Here $\rho := \sqrt{x^2 + y^2}$.
%\vskip -0.9cm
where $\rho := \sqrt{(x^1)^2 + (x^2)^2}$. 
We factor out $\rho^2$ as a \textbf{conformal factor}\index{conformal factor} to further rewrite the metric:
\begin{align}
(ds)^2(\mathbb{R}^4) = \rho^2\Big[\frac{(dx^3)^2 + (d{t})^2 + (d\rho)^2}{\rho^2} + (d\varphi)^2\Big].
%= \rho^2\left[\frac{(dx^3)^2 + (d{t})^2 + (d\rho)^2}{\rho^2} + (d\varphi)^2\right].
%\nonumber
%\label{R4toH3b}
\end{align}
Therefore, we obtain a conformal equivalence: See the left panel of Fig.1.%\ref{conformal_equiv1}.
\small
\begin{equation}
\begin{array}{cccccccc}
\mathbb{R}^4 = \mathbb{R}^3 \times \mathbb{R}^1\rightarrow&\mathbb{R}^4  &\backslash& \mathbb{R}^2 &\simeq&  \mathbb{H}^3& \times &S^1\\
&\rotatebox{90}{$\in$}&&\rotatebox{90}{$\in$}&&\rotatebox{90}{$\in$}&&\rotatebox{90}{$\in$}\\
&(x^1,x^2,x^3,{t})&&(x^3,{t})&&(\rho,x^3,{t})&&\varphi
%&(x,y,z,t)&&(z,t)&&(\rho,z,t)&&\varphi
\end{array}
%\nonumber
\end{equation}
\normalsize 
%$\bullet$ 
$\mathbb{H}^3(\rho,x^3,{t})$ is a \textbf{hyperbolic 3-space}: %with coordinates: %$(\rho, x^3,{t})$:
 $x^3,{t} \in (-\infty,+\infty)$, $\rho \in (0,\infty)$, and has the metric $g_{\mu\nu}=\rho^{-2}\delta_{\mu\nu}$ with the %\textcolor{red}
 {negative constant curvature} $-1$. 
%It is a model representing the 
This is the \textbf{upper half space model}\index{upper half space model} with $\rho>0$. 
%$\rho = 0$ corresponds to the 2-dim. plane $(x^3,{t})$.
%, but because it is a singular point.
%$\mathbb{R}^4 \backslash \mathbb{R}^2$ means excluding $(x^3,{t}) \in \mathbb{R}^2$ from $\mathbb{R}^4$.
Here $\rho = 0$ is a singularity, therefore the corresponding 2-dim. space, i.e., the $(x^3,{t})$ plane with $\rho = 0$ must be excluded from $\mathbb{R}^4$. 
%See Fig.\ref{conformal_equiv1}.
While
%\\
%$\bullet$ 
$S^1(\varphi)$ is a 1-dimensional unit sphere, i.e., a unit circle with the coordinate $\varphi \in [0,2\pi)$. 
%The special orthogonal group 
Here $SO(2)$ acts on $S^1(\varphi)$ in the standard way. 
See the left panel of Fig.~\ref{conformal_equiv_all}.

%\vskip -0.6cm
%The self-dual Yang-Mills equations are conformally invariant.
The \textbf{$SO(2) \simeq S^1$ symmetric %(or circle symmetric) 
instanton} 
solution on $\mathbb{R}^4 \backslash \mathbb{R}^2$ that does not depend on the rotation angle $\varphi$ reduces to the \textbf{hyperbolic magnetic monopole} solution on $\mathbb{H}^3$: 
the $\varphi$-rotation symmetry = $\varphi$-independence as the dimensional reduction: 
\begin{align}
x=(x^1,x^2,x^3,{t}) \equiv (\rho,%\textcolor{red}
{\varphi},x^3,{t}) \to (\rho,x^3,{t}) ,
%\nonumber
\end{align}
which is associated with the field identification %\textcolor{red}
{$\Phi(\rho,x^3,{t}) :=  \mathscr{A}_\varphi(\rho,x^3,{t})$}:
\begin{align}
 & (\mathscr{A}_{\rho}(\rho,\varphi,x^3,t), %\textcolor{red}
 {\mathscr{A}_{\varphi}}(\rho,\varphi,x^3,t), \mathscr{A}_{3}(\rho,\varphi,x^3,t), \mathscr{A}_{4}(\rho,\varphi,x^3,t)) \nonumber\\
 \to 
& (\mathscr{A}_{\rho}(\rho,x^3,{t}), %\textcolor{red}
{\Phi}(\rho,x^3,{t}), \mathscr{A}_{3}(\rho,x^3,{t}), \mathscr{A}_{4}(\rho,x^3,{t})),  \ (\rho,x^3,{t}) \in \mathbb{H}^3 .
%\nonumber
\end{align}

%\newpage
Any solution of the Bogomolny equation on $\mathbb{H}^3$ is a \textbf{$\varphi$-independent instanton solution} of the self-dual equation on $\mathbb{R}^4 \backslash \mathbb{R}^2$, ($%\textcolor{red}
{\partial_\varphi} \mathscr{A}_\ell (\rho,x^3,{t})=0$):
\begin{align}
%\mathscr{F}_{jk} = \frac{1}{\rho}\epsilon_{jk\ell}\mathscr{D}_\ell\Phi,
 (*\mathscr{F})_{\ell \varphi}(\rho,x^3,{t}) =& \frac{1}{\rho} \mathscr{D}_\ell \Phi(\rho,x^3,{t}), \ (\rho,x^3,{t}) \in \mathbb{H}^3 . %%\nonumber\\
% \Phi(\rho,x^3,{t}) :=& \mathscr{A}_\varphi(\rho,x^3,{t}) . 
%& \mathscr{F}(\rho,x^3,{t}) = *D\Phi(\rho,x^3,{t}) %\nonumber\\
%\Leftrightarrow \quad & *\mathscr{F}(\rho,x^3,{t}) = D\Phi(\rho,x^3,{t}).
%\nonumber
\end{align}
Since $S^1$ is compact (unlike $\mathbb{R}^1$), any solution of the Bogomolny equation giving a finite 3-dim. action on $\mathbb{H}^3$ gives a configuration with a finite 4-dim. action: %if $\int_{0}^{\infty}d\rho \ \rho \int_{-\infty}^{\infty}dx^3 \int_{-\infty}^{\infty}d{t} \mathscr{L}(\rho,x^3,{t}) <\infty$:
\begin{equation}
S%=&\int_{0}^{2\pi}d\varphi \int d^3x \mathscr{L}(x) %\nonumber\\ 
= \int_{0}^{2\pi}d\varphi \left[ \int_{0}^{\infty}d\rho \ \rho \int_{-\infty}^{\infty}dx^3 \int_{-\infty}^{\infty}d{t} \mathscr{L}(\rho,x^3,{t}) \right] <\infty .
%\nonumber
\end{equation}
Therefore, \textbf{$S^1\simeq SO(2)$ symmetric %(or circle symmetric) 
instantons on $\mathbb{R}^4$ can be reinterpreted as hyperbolic magnetic monopoles on $\mathbb{H}^3$, giving a configuration with a finite 4-dim. action}. 
This case (I) was first pointed out by Atiyah (1984)\cite{Atiyah84}.%, (1987)\cite{Atiyah84b}. 
%Concrete magnetic monopole solutions on $\mathbb{H}^3$ were explicitly constructed by Chakrabarti (1986)\cite{Chakrabarti86} and Nash (1986)\cite{Nash86}.

Therefore,  the hyperbolic magnetic monopoles can contribute to the path integral, because  
\begin{equation}
\exp( -S/\hbar) \not= 0 .
%\nonumber
\end{equation} 
Thus, \textbf{the hyperbolic magnetic monopoles can be responsible for quark confinement}. 

\begin{comment}
\noindent
$\odot$ 
It is possible to arbitrarily change the curvature of $\mathbb{H}^3(\rho,x^3,{t})$ by the conformal factor $\frac{\rho^2}{\rho_0^2}$:
\begin{equation}
(ds)^2(\mathbb{R}^4) = \frac{\rho^2}{\rho_0^2}\left[\rho_0^2\frac{(d{t})^2 + (dx^3)^2 + (d\rho)^2}{\rho^2} + \rho_0^2(d\varphi)^2\right] .
\label{R4toH3b}
\end{equation}
That is, the conformal equivalence is rewritten as
\begin{align}
\mathbb{R}^4 \backslash \mathbb{R}^2 \simeq \mathbb{H}^3(\rho_0) \times S^1(\rho_0) .
\end{align}
Here, $\mathbb{H}^3(\rho_0)$ denotes a %3-dim.  
hyperbolic space with negative constant curvature $-\frac{1}{2\rho_0^2}$, and $S^1(\rho_0)$ denotes a circle with radius $\rho_0$.
In the limit %of infinite radius 
$\rho_0 \to \infty$, % of $S^1(\rho_0)$, 
$S^1(\rho_0)$ approaches $\mathbb{R}$, and the curvature of $\mathbb{H}^3(\rho_0)$ approaches zero, and $\mathbb{H}^3(\rho_0)$ reduces to %the flat Euclidean space
 $\mathbb{R}^3$.

\end{comment}

%%%%%%%%%%%%%%%%%%%%%%%%%%%%%%%%%%%%%%%%%%%%%%%%%%%%%%%%%%%
%%%%%%%%%%%%%%%%%%%%%%%%%%%%%%%%%%%%%%%%%%%%%%%%%%%%%%%%%%%

%\newpage
\section{Conformal equivalence (II)}
%\noindent
%\textcolor{green}{\Large\bf $\S$~~Conformal equivalence (II)
%}
%\setcounter{equation}{0}
\normalsize 
%\small
%Let us consider another example.% of conformal equivalence.
%\\
%$\odot$ \noindent
(II)
We consider another solution with \textbf{spatial rotation symmetry}: $SO(3)$.
%\begin{example}[2-dimensional hyperbolic space $\mathbb{H}^2$]
%The metric of $\mathbb{R}^4$ is given by using the Cartesian coordinates $({t},x^1,x^2,x^3)$:
%\begin{align}
%(ds)^2(\mathbb{R}^4) = (d{t})^2 + (dx^1)^2 +(dx^2)^2 +(dx^3)^2 .
%\end{align}
\\
We introduce the polar coordinates $(r,\theta,\varphi)$ for the 3-dim. space $(x^1,x^2,x^3)$: %to rewrite the metric as 
\begin{align}
(ds)^2(\mathbb{R}^4) =  (d{t})^2 + (dr)^2 +r^2((d\theta)^2 +\sin^2\theta(d\varphi)^2) , %\ (r := \sqrt{x_1^2 + x_2^2 + x_3^2}) .
%\nonumber
\end{align}
where $r := \sqrt{(x^1)^2 + (x^2)^2 + (x^3)^2}$.
Then, we factor out $r^2$ as a \textbf{conformal factor}\index{conformal factor} to rewrite %the metric 
\begin{equation}
(ds)^2(\mathbb{R}^4) =r^2 \Big[ \frac{(d{t})^2 + (dr)^2}{r^2} +((d\theta)^2 +\sin^2\theta(d\varphi)^2) \Big].
%\left[ \frac{(d{t})^2 + (dr)^2}{r^2} +(d\theta)^2 +\sin^2\theta(d\varphi)^2 \right] .
%\nonumber
%\label{R4toH2}
\end{equation}
Therefore, we obtain the \textbf{conformal equivalence}\index{conformal equivalence}: See the right panel of Fig.1.%\ref{conformal_equiv2}.
\begin{align}
%\footnotesize
\begin{array}{cccccccc}
\mathbb{R}^4 = \mathbb{R}^2 \times \mathbb{R}^2\rightarrow&\mathbb{R}^4  &\backslash& \mathbb{R}^1 &\simeq&  \mathbb{H}^2&\times &S^2\\
&\rotatebox{90}{$\in$}&&\rotatebox{90}{$\in$}&&\rotatebox{90}{$\in$}&&\rotatebox{90}{$\in$}\\
&(t,x,y,z)&&t&&(t,r)&&(\theta,\varphi)
\end{array}
%\nonumber
\end{align}
%\vskip -0.6cm

%\noindent
%$\bullet$ 
$\mathbb{H}^2({t},r)$ is a \textbf{hyperbolic plane}\index{hyperbolic plane} with %coordinates $({t},r)$: 
${t} \in (-\infty,\infty)$, $r \in(0,\infty)$, and has the metric $g_{\mu\nu}=r^{-2}\delta_{\mu\nu}$ with the %\textcolor{red}
{negative constant curvature} $(-1)$. 
This is the \textbf{upper half plane model}\index{upper half plane model} with $r >0$. Here $r=0$ is a singularity: %, therefore %the one-dimensional space $\mathbb{R}^1$ with $r = 0$, i.e., 
the ${t}$-axis must be excluded from $\mathbb{R}^4$.
%See Fig.\ref{conformal_equiv2}.
While
%\\
%$\bullet$ 
$S^2(\theta,\varphi)$ is a two-dimensional unit sphere with %coordinates $(\theta, \varphi)$: 
$\theta \in [0,\pi)$, $\varphi \in [0,2\pi)$ and has a positive constant curvature (2). %(Unit means radius is 1)
%\footnote{
%N. Manton and P. Sutcliffe (2004), \S 4.3, \S 10.1, p.424.
%}
%The special orthogonal 
$SO(3)$ acts on $S^2(\theta,\varphi)$ in the standard way.
%\vskip -0.9cm
%\noindent
The \textbf{$SO(3)$ (spherically) symmetric instanton} on $\mathbb{R}^4 \backslash \mathbb{R}^1$ that does not depend on the rotation angles $\theta,\varphi$ reduces to  \textbf{hyperbolic vortex} on $\mathbb{H}^2(r,{t})$.
See the right panel of Fig.~\ref{conformal_equiv_all}.

Any  solution of the vortex equation on $\mathbb{H}^2(r,{t})$ is a $\theta,\varphi$-independent solution of  self-dual equation on $\mathbb{R}^4 \backslash \mathbb{R}^1$
 which is written 
for $a_t=a_t(r,{t}) , a_r=a_r(r,{t}) , \ \phi_1=\phi_1(r,{t}) ,  \ \phi_2=\phi_2(r,{t}), \ (r,{t}) \in \mathbb{H}^2$:
\begin{align}
\left\{\,
\begin{aligned}
& \partial_{t}a_r - \partial_ra_t = \frac{1}{r^2}(1 - \phi_1^2 - \phi_2^2), \\
&\partial_{t}\phi_1 + a_t\phi_2 = \partial_r\phi_2 - a_r\phi_1, \
 \partial_{t}\phi_2 - a_t\phi_1 = -(\partial_r\phi_1 + a_r\phi_2) .
\end{aligned} 
\right.
%\nonumber
%\label{vortex_eq1}
%\mathscr{L}=\mathscr{L}(r,{t})
%\mathscr{F}(r,{t}) = *D\Phi(r,{t}) 
%\Leftrightarrow  *\mathscr{F}(r,{t}) = D\Phi(r,{t}).
\end{align}
Any solution of the %\textcolor{red}
{vortex equation} giving a finite 2-dim. action on $\mathbb{H}^2(r,{t})$ $\int_{0}^{\infty}dr \ r^2 \int_{-\infty}^{\infty}d{t} \mathscr{L}(r,{t}) <\infty$  gives %a configuration with 
a finite 4-dim. action: 
$%\begin{equation}
S%=&\int_{0}^{\pi}d\theta \int_{0}^{2\pi}d\varphi \int d^2x \mathscr{L}(x) %\nonumber\\ 
= \int_{0}^{\pi}d\theta \sin \theta \int_{0}^{2\pi}d\varphi \left[ \int_{0}^{\infty}dr \ r^2 \int_{-\infty}^{\infty}d{t} \mathscr{L}(r,{t}) \right] <\infty ,
$ %\end{equation}
%if $\int_{0}^{\infty}dr \ r^2 \int_{-\infty}^{\infty}d{t} \mathscr{L}(r,{t}) <\infty$.
since $S^2(\theta,\varphi)$ is compact.  % (unlike $\mathbb{R}^2$). 

Therefore, $SO(3)$ \textbf{spherically symmetric instantons on $\mathbb{R}^4$ can be reinterpreted as vortices on $\mathbb{H}^2$, giving a configuration with a finite 4-dim.  action}.
This case (II) was discovered by Witten (1977)\cite{Witten}. %(681 citations)
 to find multi-instanton solutions of 4-dim.  Yang-Mills theory, which is %confirmed by %Manton(1978).%\cite{Manton78}.
%This view was 
established  as the symmetric instanton by 
Forgacs and Manton (1980)\cite{FM80}. % (594 citations).
%It is also written in Manton and Sutcliffe, p.424, section 4.3 (2004)\cite{MS04}.
%\\
Therefore,  the hyperbolic vortices can contribute to the path integral due to  $\exp( -S/\hbar) \not= 0$ and \textbf{the hyperbolic vortices can be responsible for quark confinement}.

\begin{figure}[htb]
\begin{center}
\includegraphics[scale=0.5]{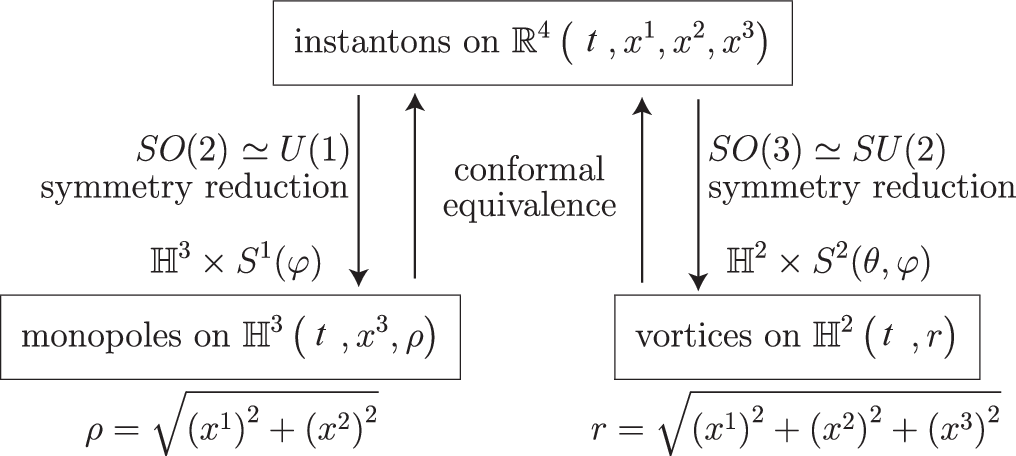}
\end{center}
\caption{
Symmetric instanton, conformal equivalence, and dimensional reduction.
}
\label{conformal_unification}
\end{figure}

The results (I), (II) are summarized in Fig.~\ref{conformal_unification}.
%\textcolor{blue}{
Conformal equivalence reshapes the background geometry, while symmetry reduction eliminates dependence of field content on compact directions.
%\\
%conformal equivalence→ background geometry changes
%\\
%dimensional reduction→ field content reduces
%}
%\textcolor{blue}{
The crucial point is that translation-invariant monopoles in flat 
$\mathbb{R}$ have infinite four-dimensional action, whereas rotation-invariant instantons effectively compactify the reduced directions, rendering the action finite.
%}
%\textcolor{blue}{
In the semi-classical regime, these finite-action configurations dominate the infrared path integral and thus control the Wilson loop behavior.
%}

%%%%%%%%%%%%%%%%%%%%%%%%%%%%%%%%%%%%%%%%%%%%%%%%%%%%%%%%%%%
%%%%%%%%%%%%%%%%%%%%%%%%%%%%%%%%%%%%%%%%%%%%%%%%%%%%%%%%%%%

%\newpage
\section{Unifying magnetic monopole and vortices}
\noindent
\textbf{Definition} [Rotationally symmetric gauge field](Manton and Sutcliffe(2004)\cite{MS04})
\\
%When a gauge field $\mathscr{A}(x)$ with a non-Abelian gauge group $G$ undergoes a spatial rotation $R$ in $\mathbb{R}^d(d>2)$, and the gauge field can be made invariant by simultaneously performing an appropriate gauge transformation $U_R(\bm{x})$, the gauge field $\mathscr{A}(x)$ is called \textbf{rotationally symmetric}\index{rotationally symmetric} gauge field. 
%We say that the rotation keeps the field invariant, apart from the gauge transformation. 
%When a space rotation $R$ is performed on the spatial component $\mathscr{A}_j(\bm{x})$ of the gauge field, if the corresponding gauge transformation $U_R(\bm{x})$ satisfies
If the space rotation $R$ has the same effect on the gauge field as the gauge transformation $U_R$: 
\begin{align}
R_{kj}\mathscr{A}_k(R\bm{x}) =& U_R(\bm{x})\mathscr{A}_j(\bm{x})U_R^{-1}(\bm{x}) + iU_R(\bm{x})\partial_jU_R^{-1}(\bm{x}) ,
%\nonumber
%\label{eq:5}
\end{align}
the gauge field $\mathscr{A}(x)$ is called %\textcolor{red}
\textit{rotationally symmetric}. 
%then $\mathscr{A}_j$ is invariant under the space rotation. 
%In other words, the space rotation $R$ has the same effect on the gauge field as the gauge transformation $U_R$. 
Or equivalently, if we combine $R$ and $U_R^{-1}$, the gauge field remains invariant.
\begin{comment}
\\
If the gauge field $\mathscr{A}$ is combined with a scalar field $\Phi$ that transforms as a fundamental representation of $G$ under a gauge transformation of the group $G$, if the relation holds:
\begin{equation}
\Phi(R\bm{x}) = U_R(\bm{x})\Phi(\bm{x}) , 
%\nonumber
%\label{eq:6}
\end{equation}
$\Phi$ is invariant under rotations.
\end{comment}
%\end{definition}

%\newpage
%\begin{longshadowbox}%\begin{breakbox}%\begin{itembox}[c]{}
%\begin{prop}
\noindent
\textbf{Proposition}%[Witten分解]
%\noindent
%$\odot$ 
[Witten transformation (Witten Ansatz) for %\textcolor{red}
{$SO(3)$ symmetric gauge field}]
\\
The $D=4$ $SU(2)$ Yang-Mills field $\mathscr{A}_\mu(x)$ with $SO(3)$ spatial rotation symmetry on $\mathbb{R}^4(x)$ is dimensionally reduced to the $D=2$ gauge field $a_t(r,{t}), a_r(r,{t})$ and the scalar field $\phi_1(r,{t}), \phi_2(r,{t})$ on $\mathbb{H}^2(t,r)$ through the transformation which we call the %\textcolor{red}
\textit{Witten transformation} (called the Witten Ansatz):
%The transformation with the $SO(3)$ spatial rotation symmetry from the $D=4$ $SU(2)$ Yang-Mills field to the dimensionally reduced $D=2$ field is given by the \textcolor{red}{Witten transformation} (which was originally called the Witten Ansatz):
\begin{align}
\mathscr{A}_{t}(x) = \frac{\sigma_A}{2}\frac{x^A}{r}a_t(r,{t}) , %\ [r:=\sqrt{(x^1)^2+(x^2)^2+(x^3)^2} , \ (r,{t}) \in \mathbb{H}^2] .
%\nonumber\\
\quad
\mathscr{A}_j(x) =\frac{\sigma_A}{2}\left\{\frac{x^A}{r}\frac{x^j}{r}a_r(r,{t}) + \frac{\delta_j^Ar^2 - x^Ax^j}{r^3}\phi_1(r,{t}) + \epsilon_{jAk}\frac{x^k}{r^2}[1 + \phi_2(r,{t})]\right\}.
%\nonumber
\end{align}
%The Witten transformation is the most general form (which is unique up to the gauge transformation) for the field that is invariant under a combination of $SO(3)$ spatial rotation (spacetime symmetry) and $SU(2)$ gauge transformation (internal symmetry).
%\footnote{
%E. Witten (1979).
%The dimensionally reduced field has 4-dim.  cylindrical symmetry, i.e., SO(3) rotational symmetry around the t-axis.

%The hyperbolic vortex on $\mathbb{H}^2$ consists of a complex scalar field $\phi = \phi_1(r,{t}) + i\phi_2(r,{t})$ and a gauge potential $a = a_r(r,{t})dr + a_t(r,{t})d{t}$. These hyperbolic vortex fields are functions of ${t}$ and $r$.

%\newpage
%\begin{longshadowbox}%\begin{breakbox}%\begin{itembox}[c]{}
%\begin{prop}
\noindent
\textbf{Proposition} [magnetic monopole on $\mathbb{H}^3$,  vortex on $\mathbb{H}^2$]\label{prop:monopole-vortex}
%In the polar coordinate $(\rho, \varphi, x^3, {t})$,
%we define the $\rho$-component $\mathscr{A}_\rho$ and the $\varphi$-component $\mathscr{A}_\varphi$ of the Yang-Mills field $\mathscr{A}$ by
%\begin{align}
%\mathscr{A}_\rho(x):=\frac{1}{\rho}(x^1\mathscr{A}_1(x) + x^2\mathscr{A}_2(x)) , \quad 
%\mathscr{A}_{\varphi}(x):= -x^2\mathscr{A}_{1}(x) +x^1\mathscr{A}_{2}(x) .
%%\nonumber
%\end{align}
%以上より$\mathscr{A}_\rho$の次式を得る。
%\begin{align}
%\fbox{$
%\begin{aligned}
%\mathscr{A}_\rho =& \frac{1}{2}\left\{\sigma_A n^A \frac{\rho}{r}a_r + \frac{x_3}{r^3}(\sigma_1x_3\cos\varphi +\sigma_2x_3\sin\varphi - \sigma_3\rho)\phi_1 \right. %\nonumber\\
%&\quad \left. +\frac{x_3}{r^3}(-\sigma_1\sin\varphi +\sigma_2\cos\varphi)(1+\phi_2)\right\}.
%\end{aligned}$}
%\end{align}
%一方，$\mathscr{A}$の$\varphi$成分$\mathscr{A}_\varphi$を定義する：
%\begin{align}
%\boxed{\mathscr{A}_{\varphi}(x):= -x^2\mathscr{A}_{1}(x) +x^1\mathscr{A}_{2}(x)} .
%\end{align}
%以上より$\mathscr{A}_{\varphi}$の次式を得る：
%\begin{align}
%\fbox{$
%\begin{aligned}
%\mathscr{A}_{\varphi} =& \frac{1}{2}\left\{\frac{\rho}{r}(-\sigma_1\sin\varphi + \sigma_2\cos\varphi)\phi_1 \right. %\nonumber\\
%&\quad \left. +\frac{\rho}{r^2}(-\sigma_1x_3\cos\varphi -\sigma_2x_3\sin\varphi+\sigma_3\rho)(1+\phi_2)\right\}.
%\end{aligned}$}
%\end{align}
%\end{prop}
%\end{longshadowbox}%\end{breakbox}%\end{itembox}
We can apply the gauge transformation 
\begin{equation}
U_\varphi=\exp \left( i\varphi\frac{\sigma_3}{2} \right) \in SU(2)  \ \left( \varphi:=\arctan \frac{x^2}{x^1} \in [0, 2\pi) \right)
%\nonumber
\end{equation} 
corresponding to a rotation around the $x_3$ axis by an angle $\varphi$  
to the instanton: %$\mathscr{A}_\mu(x^1,x^2,x^3,{t})$ 
\begin{align}
  \mathscr{A}_\mu(x^1,x^2, x^3,{t})  
\rightarrow    U_\varphi \mathscr{A}_\mu(x^1,x^2, x^3,{t}) U_\varphi^\dagger + iU_\varphi\partial_\mu U_\varphi^\dagger
  =:  \mathscr{A}_\mu^G(\rho, x^3,{t}) .
%%\nonumber
\end{align}
so that $\mathscr{A}_\mu(x^1,x^2,x^3,{t})$ becomes independent of $\varphi$, leading to an $S^1$-symmetric instanton $\mathscr{A}_\mu^G(\rho,x^3,{t})$. 
% ($\rho := \sqrt{(x^1)^2+(x^2)^2}$).
%In this case, $\mathscr{A}_\varphi^G(\rho,x^3,{t})$ is identified with the hyperbolic magnetic monopole field $\Phi(\rho,x^3,{t})$ on $\mathbb{H}^3$: 
%\begin{align}
%\mathscr{A}_\varphi^G(\rho,x^3,{t}) = \Phi(\rho,x^3,{t}).
%%\nonumber
%\end{align}
%The gauge field  
%$\mathscr{A}_\rho^G(\rho,x^3,{t})$, $\mathscr{A}_3^G(\rho,x^3,{t})$, $\mathscr{A}_4^G(\rho,x^3,{t})$ and %the scalar field $\Phi(\rho,x^3,{t})$ of a 
Then the magnetic monopole on $\mathbb{H}^3(\rho,x^3,{t})$ is written in terms of %the gauge field
%$a_t=a_t(t,r)$, $a_r=a_r(t,r)$,  %the scalar field
%$\phi_1=\phi_1(t,r)$, $\phi_2=\phi_2(t,r)$ of 
the vortex on $\mathbb{H}^2(r,{t})$: 
\begin{align}
\mathscr{A}_t^G(\rho, x^3,{t}) =&\frac{1}{2}\left\{\frac{1}{r}(\sigma_1\rho + \sigma_3x_3)\right\}a_t(r,{t}),  %\ (\rho, x^3,{t}) \in \mathbb{H}^3 ,
\nonumber
%\label{eq:4}
\\
\mathscr{A}_3^G(\rho, x^3,{t}) =&\frac{1}{2}\left\{\frac{x^3}{r^2}(\sigma_1\rho + \sigma_3x^3)a_r(r,{t}) + \frac{\rho}{r^3}(-\sigma_1x^3 +\sigma_3\rho)\phi_1(r,{t}) - \frac{\rho}{r^2}\sigma_2(1 + \phi_2(r,{t}))\right\}, 
\nonumber
%\label{eq:5}
\\
\mathscr{A}_\rho^G(\rho, x^3,{t}) =&\frac{1}{2}\left\{\frac{\rho}{r^2}(\sigma_1\rho + \sigma_3x^3)a_r(r,{t}) + \frac{x^3}{r^3}(\sigma_1x^3 -\sigma_3\rho)\phi_1(r,{t}) + \frac{x^3}{r^2}\sigma_2(1 + \phi_2(r,{t}))\right\}, 
\nonumber
%\label{eq:6}
\\
\Phi(\rho, x^3,{t}) %\equiv \mathscr{A}_\varphi^G(\rho, x^3,{t}) 
=& \frac{1}{2}\left\{\frac{\rho}{r}\sigma_2\phi_1(r,{t}) + \frac{\rho}{r^2}(-\sigma_1x^3 +\sigma_3\rho)(1 + \phi_2(r,{t})) +\sigma_3\right\}.
%\nonumber
%\label{eq:7}
\end{align}
%where $\sigma_A (A=1,2,3)$ are the Pauli matrices and 
%$\rho := \sqrt{(x^1)^2+(x^2)^2}$, $r:= \sqrt{(x^1)^2+(x^2)^2+(x^3)^2}=\sqrt{\rho^2+(x^3)^2}$.
%
%\end{prop}
%\end{longshadowbox}%\end{breakbox}%\end{itembox}
%
%\begin{proof}
Although this result was obtained by Maldonado (2017)\cite{Maldonado17}, 
%by straightforward but a bit tedious calculations. 
%Here the assignment of the components are different from his result for our  convenience. 
it is modified for our later convenience. 
%\end{proof}
%\textcolor{blue}{
Equation (24) provides a non-trivial explicit map between monopole and vortex fields, going beyond qualitative correspondence.
%}
%
%\begin{align}
% a_t=a_t(t,r) , \ \phi_1=\phi_1(t,r) ,  \ \phi_2=\phi_2(t,r), \ (t,r) \in \mathbb{H}^2 ,
%\end{align}
%$a_t=a_t(t,r)$, $\phi_1=\phi_1(t,r)$, $\phi_2=\phi_2(t,r)$, 
%$r:= \sqrt{(x^1)^2+(x^2)^2+(x^3)^2}=\sqrt{\rho^2+(x^3)^2}$および$\rho := \sqrt{(x^1)^2+(x^2)^2}$である。
%\end{prop}
%\end{longshadowbox}%\end{breakbox}%\end{itembox}
%\begin{comment}
%\newpage
%\noindent
%$\odot$ 
The relationship for the norm between the $su(2)$-valued hyperbolic magnetic monopole field $\Phi(\rho, x^3,{t})=\mathscr{A}_\varphi^G(\rho, x^3,{t})$ and the complex-valued hyperbolic vortex field $\phi({t},r)=\phi_1({t},r)+i\phi_2({t},r)$ is given as 
\begin{align}
 ||\Phi({t},x^3,\rho)||^2 = \frac{\rho^2|\phi({t},r)|^2 + (x^3)^2}{tr^2} , \quad  r:=\sqrt{\rho^2 + (x^3)^2}  .
%\nonumber
\end{align}
%$\mathbb{H}^3$ is the upper half-space $\rho>0$ of $({t},x^3,\rho)$, and its boundary $\partial\mathbb{H}^3$ is the plane $\partial\mathbb{H}^3 = ({t},x^3)$. 
The norm $||\Phi||$ has the correct boundary value: $||\Phi|| \rightarrow v = \frac{1}{2}$ $(\rho\rightarrow0)$.
\section{Holography: bulk/boundary correspondence}
%\noindent
%\textcolor{green}{\Large\bf $\S$~Holography: bulk/boundary correspondence}
%\setcounter{equation}{0}
%\normalsize 
%\small
%\\
It was rigorously shown 
%\textcolor{blue}{
%rigorous in the sense of uniqueness modulo gauge equivalence
%}
that \textbf{the %\textcolor{red}
\textit{holographic principle} 
%(`t Hooft (1993)%\cite{Hooft93} Susskind (1995)%\cite{Susskind95})
applies to %\textcolor{red}
{hyperbolic magnetic monopoles} in the hyperbolic space $\mathbb{H}^3$}. 
In contrast, %it should be notted that 
\textbf{it does not apply to magnetic monopoles in flat Euclidean space $\mathbb{R}^3$}. 
See references in \cite{Kondo25}.
%, even though they are regarded as the infinite mass limit of  hyperbolic magnetic monopoles in $\mathbb{H}^3$}. 
%In flat space, the holographic images of any two magnetic monopoles with the same magnetic charge on $S^2$ are identical and indistinguishable.
%(`t Hooft (1993)%\cite{Hooft93}
%, Susskind (1995)%\cite{Susskind95}
%)
%\textcolor{blue}{
Here `holography' means the uniqueness of hyperbolic monopole solutions given their asymptotic boundary data, modulo gauge equivalence.
%}
%\begin{longshadowbox}
%\begin{prop}
\\
\noindent
\textbf{Proposition} [Bulk/boundary correspondence of $\mathbb{H}^3=AdS_3$]
A magnetic monopole on hyperbolic space (anti-de Sitter space) $\mathbb{H}^3=AdS_3$ is completely determined by its asymptotic boundary value %(the value of the boundary 
at infinity $\partial\mathbb{H}^3$, apart from the gauge equivalence. 
This situation is in sharp contrast with the Euclidean case in which all monopole have the same boundary values. 
\noindent
\textbf{Proposition} [%\textcolor{red}
{Abelian dominance and magnetic monopole dominance} on  $\partial\mathbb{H}^3$]
On the conformal boundary $\partial\mathbb{H}^3 \simeq S^2$ of %the upper half-space coordinates of  
$\mathbb{H}^3(\rho,x^3,{t})$, that is, $\rho \to 0$: ${t}$-$x^3$ plane, the $SU(2)$ Yang-Mills field and the $SU(2)$ scalar field converges to
\begin{align}
\mathscr{A}_{t}^G(\rho,x^3,{t}) &%= \frac{\sigma_3}{2}\frac{x^3}{r}a_t 
\to \frac{%\textcolor{red}
{\sigma_3}}{2}a_t({t},x^3) , \ %%\nonumber\\
\mathscr{A}_3^G(\rho,x^3,{t})  %= \frac{\sigma_3}{2}\frac{(x^3)^2}{r^2}a_r  
\to \frac{%\textcolor{red}
{\sigma_3}}{2}a_r({t},x^3) , \nonumber\\
\mathscr{A}_\rho^G(\rho,x^3,{t}) &%= \frac{\sigma_1}{2}\frac{(x^3)^2}{r^3}\phi_1 + \frac{\sigma_2}{2}\frac{x^3}{r^2}(1 + \phi_2) 
\to \frac{%\textcolor{blue}
{\sigma_1}}{2}\frac{1}{r}\phi_1({t},x^3) + \frac{%\textcolor{blue}
{\sigma_2}}{2}\frac{1}{r} [1 + \phi_2({t},x^3)], \nonumber\\
\Phi(\rho,x^3,{t}) &\to \frac{%\textcolor{red}
{\sigma_3}}{2}(1) \ \left( ||\Phi || \to v=\frac12 \right) .
%\nonumber
\end{align}
%where $r = \sqrt{\rho^2 + (x^3)^2} \to |x^3|$ for $\rho \to 0$. 
Therefore, 
the gauge field $\mathscr{A}_\rho^G(\rho,x^3,{t})$ %\textcolor{blue}
{in the bulk direction}  is dominated by the %\textcolor{blue}
{off-diagonal components}, % $\frac{\sigma_1}{2}\frac{1}{r}\phi_1({t},x^3) + \frac{\sigma_2}{2}\frac{1}{r} [1 + \phi_2({t},x^3)]$.
while the gauge field $\mathscr{A}_4^G(\rho,x^3,{t}), \mathscr{A}_3^G(\rho,x^3,{t})$ %\textcolor{red}
{on the boundary} $\rho=0$ has only the %\textcolor{red}
{diagonal components} $a_t({t},x^3), a_r({t},x^3)$.

\section{Quark confinement: area law of Wilson loop average}
%\noindent
%\textcolor{green}{\Large\bf $\S$~Quark confinement: area law of Wilson loop average
%}
%\setcounter{equation}{0}
%\normalsize 
%\small
%Finally, we define the Wilson loop operator for a closed loop $C$, and calculate its expectation value using the dilute gas approximation of hyperbolic magnetic monopoles and hyperbolic vortices to show the area law.  
%Therefore, confinement is understood in the sense of a linear potential for quark--anti-quark static potential.
%
%\newpage
%\begin{definition}
\noindent
\textbf{Definition}  [Wilson loop operator]
Let $\mathscr{A}$ be a Lie algebra valued \textbf{connection 1-form}:
\begin{equation}
\mathscr{A}(x) := \mathscr{A}_\mu(x) dx^\mu = \mathscr{A}_\mu^A(x) T_A dx^\mu.
\end{equation}
For a given loop $C$, the \textbf{Wilson loop operator} $W_{\rm C}[\mathscr{A}]$ in the representation $\mathcal{R}$ is defined using the \textbf{path ordered product} $\mathscr{P}$: 
\begin{equation}
W_C[\mathscr{A}] := %\mathcal{N}^{-1} 
{\rm tr}_{\mathcal{R}} \left\{ \mathscr{P} \exp \left[ ig \oint_C \mathscr{A} \right] \right\} /{\rm tr}_{\mathcal{R}}(1).
\end{equation}
%where $\mathscr{P}$ represents the \textbf{path ordered product}.%, and the normalization factor $\mathcal{N}$ is equal to the dimension $d_R$ of the representation $R$ to which the Wilson loop probe belongs.%, leading to $W_C[0]=1$.
%The Yang-Mills coupling constant $g_{{}_{\rm YM}}$ 
%is introduced for later convenience, but it 
%can be eliminated by scaling the field: $\mathscr{A}  \to g_{{}_{\rm YM}}^{-1}\mathscr{A}$.

\begin{figure}[htb]
\begin{center}
\includegraphics[scale=0.35]{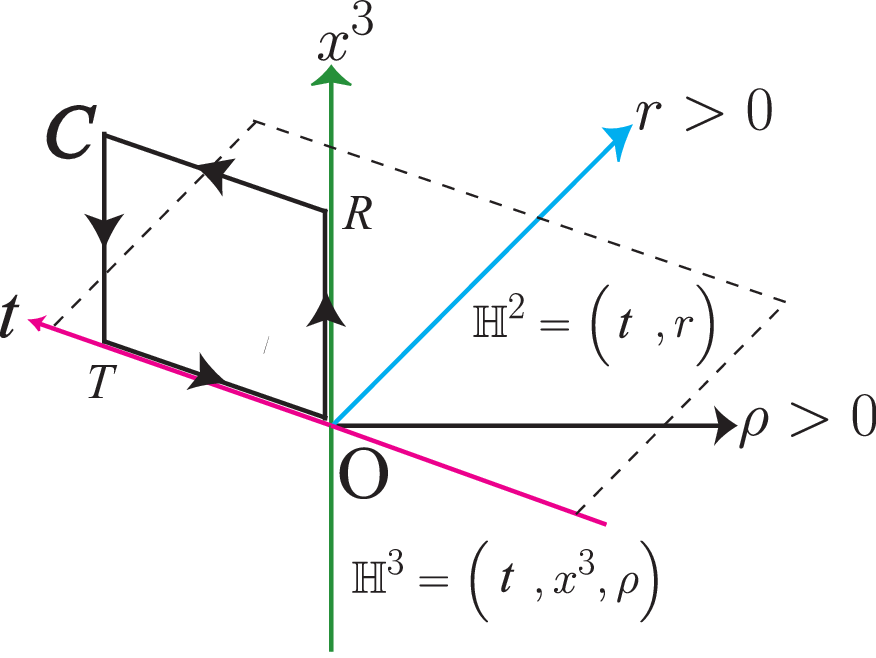}
\quad
\includegraphics[scale=0.35]{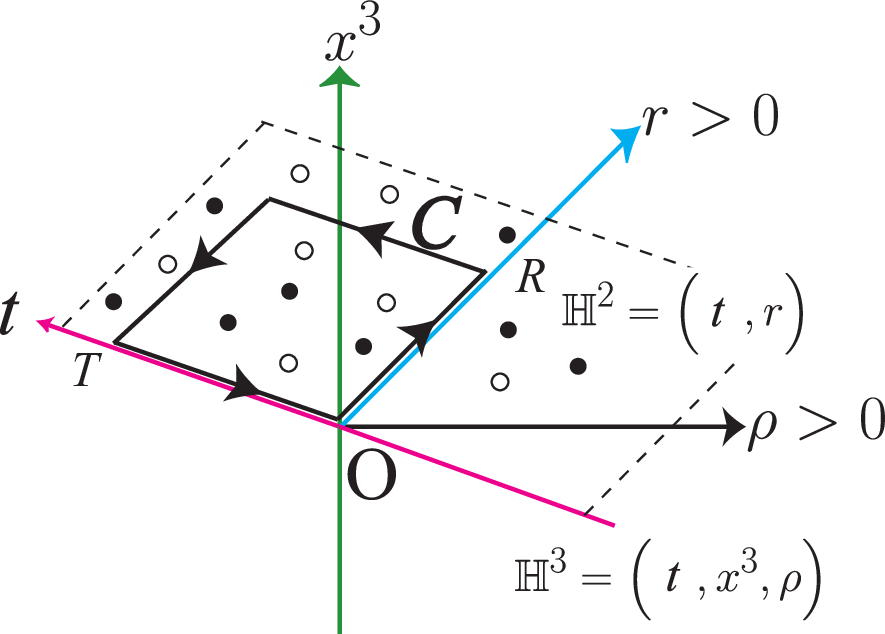}
\end{center}
\caption{
(Left) 
The Wilson loop $C$ located on the conformal boundary $\partial \mathbb{H}^3$, i.e., $t-x^3$ plane of the hyperbolic space $\mathbb{H}^3$.
(Right) 
The Wilson loop $C$ located on $\partial \mathbb{H}^2$, i.e., $t-r$ plane of the hyperbolic plane $\mathbb{H}^2$.
Hyperbolic vortices (black circles) and anti-vortices (white circles)  located inside and outside of the Wilson loop $C$ on $\mathbb{H}^2$.%  in the dilute gas picture.
% and the hyperbolic magnetic monopole (white circle) on $\mathbb{H}^3$.
%(Right) The dilute gas approximation. 
}
\label{Wilson_vortex_monopole}
\end{figure}

%\end{definition}
\noindent
(I) Quark confinement due to %\textcolor{red}
{hyperbolic magnetic monopoles} on $\mathbb{H}^3$ and %\textcolor{red}
{holography}: 
%\\
%Next, we calculate the the Wilson loop average in a different setting. 
We locate the Wilson loop $C$ on the boundary $\partial\mathbb{H}^3(x^3,{t})$ of $\mathbb{H}^3$ in the limit $\rho \to 0$.
See the left panel of Fig.~\ref{Wilson_vortex_monopole}. 
 
%\begin{longshadowbox}%\begin{breakbox}%\begin{itembox}[c]{}
%\begin{prop}
\noindent
\textbf{Proposition} [Wilson loop operator on the conformal boundary $\partial \mathbb{H}^3$]
If the loop $C$ lies on the conformal boundary $\partial \mathbb{H}^3$, i.e., $x^3-{t}$ of $\mathbb{H}^3$, the Wilson loop operator in the fundamental representation $F$ defined for the $S^1$-invariant $SU(2)$ Yang-Mills field $\mathscr{A}_\mu^G$ takes the simple Abelian form as $\rho \to 0$:
\begin{align}
W_C[\mathscr{A}] 
= \frac12  {\rm tr}_{F} \left\{ \exp \left[ i \frac{\sigma_3}{2}  \oint_C dx^\mu a_\mu({t},x^3) \right] \right\} %%\nonumber\\
= \frac12  {\rm tr}_{F} \left\{ \exp \left[ i \frac{\sigma_3}{2}  \int_{\Sigma: \partial \Sigma=C} dt dx^3  F_{tr}({t},x^3) \right] \right\} 
.
%\nonumber
\end{align}
%This Wilson loop operator is the same as the previous one (\ref{Wilson_loop_A}). 
%Adopting the dilute instanton gas approximation, the Wilson loop average obeys the area law.% (\ref{wler}). 
%\end{prop}
%\end{prop}
%\end{longshadowbox}%\end{breakbox}%\end{itembox}
The $SU(2)$ field strength on the boundary has only the maximal torus  $U(1)$ component: 
\begin{align}
\mathscr{F}_{t3}^G(\rho,x^3,{t}) %:=& \partial_{t}\mathscr{A}_3^G - \partial_3\mathscr{A}_4^G - ig[\mathscr{A}_4^G, \mathscr{A}_3^G] %%\nonumber\\
\to %\frac{\sigma_3}{2}\frac{(x^3)^2}{r^2}\partial_{t}a_r - \frac{\sigma_3}{2}\partial_3\left(\frac{x^3}{r}a_t\right) %%\nonumber\\
%=  
\frac{\sigma_3}{2}(\partial_{t}a_r - \partial_ra_t) 
=\frac{\sigma_3}{2} F_{tr}({t},x^3).
\end{align}
Therefore, the Yang-Mills field reduces to the diagonal Abelian field on the conformal boundary $x^3-{t}$.  
This fact is regarded as the (infrared) %\textcolor{red}
\textbf{Abelian dominance} and the %\textcolor{red}
\textbf{magnetic monopole dominance} in quark confinement.
%,  which is expected but not proved in the Euclidean case. 
%This is regarded as the \textbf{Abelian dominance}  and the \textbf{magnetic monopole dominance} in quark confinement.
In the ordinary flat Euclidean case,  (infrared) Abelian dominance and  magnetic monopole dominance in quark confinement have been confirmed by numerical simulations and also supported by analytical investigations, but not proved rigorously in the Euclidean case.
See e.g., \cite{PR}.%Kondo, Kato, Shibata and Shinohara, Phys.Rept.\textbf{579}, 1--226 (2015), e-Print: 1409.1599 [hep-th]%\cite{PR} for the details, although there are no rigorous proofs. 

%%%%%%%%%%%%%%%%%%%%%%%%%%%%%%%%%%%%%%%%%%%%%%%%%%%%%%%%%%%
%%%%%%%%%%%%%%%%%%%%%%%%%%%%%%%%%%%%%%%%%%%%%%%%%%%%%

\begin{figure}[htb]
\begin{center}
\includegraphics[scale=0.3]{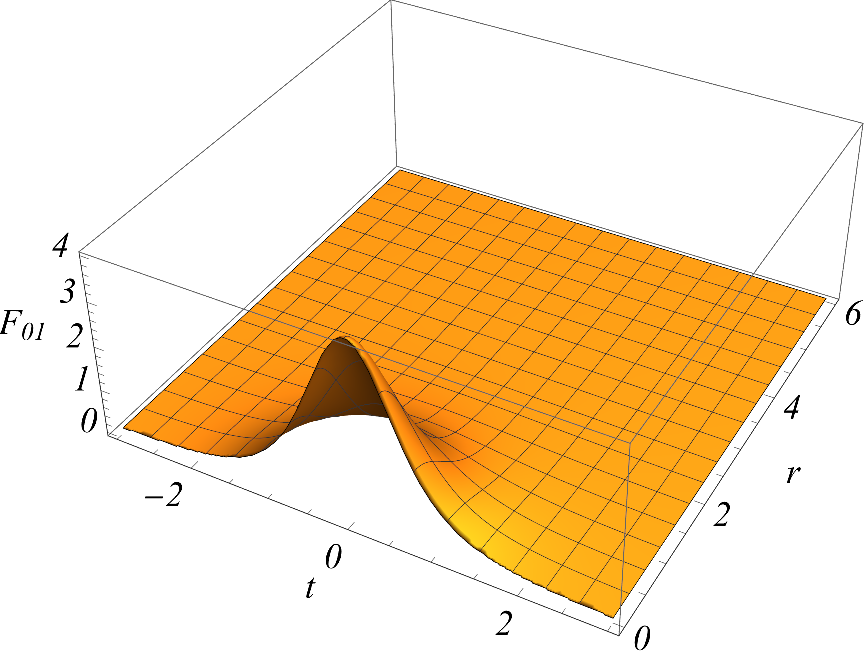}
\quad\quad
\includegraphics[scale=0.3]{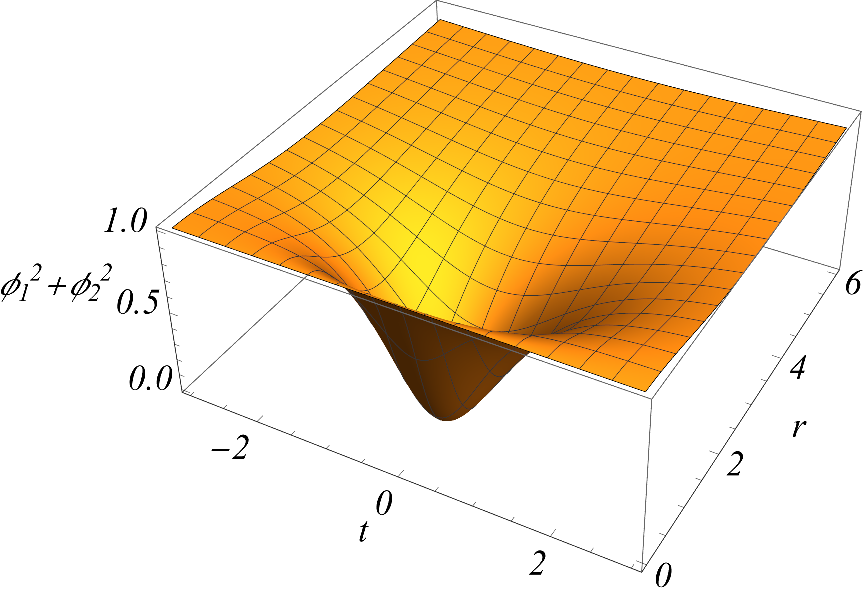}
\end{center}
%\vskip -0.2cm
\caption{
The 1-vortex solution with the center at $(t,r)=(0,1)$ and the size $\lambda=1$. The distribution of gauge-invariant quantities . 
(Left) field strength $F_{tr}(t,r)$, (Right) $|\phi(t,r)|^2$. 
}
\label{F_vortex}
\end{figure}

%\newpage
%\begin{longshadowbox}%\begin{breakbox}%\begin{itembox}[c]{}
%\begin{prop}
\noindent
(II) Quark confinement due to %\textcolor{red}
{hyperbolic vortices} on $\mathbb{H}^2$: 
%We locate the Wilson loop $C$ on the $(t,r)$ plane on $\mathbb{H}^2$. 
See the right panel of Fig.~\ref{Wilson_vortex_monopole}. 
%\textcolor{blue}{
We can use the geometric picture that vortices and anti-vortices puncture the minimal surface bounded by C to evaluate the Wilson loop average by counting the intersection numbers in the dilute gas approximation. 
%}
%\begin{comment}
The vortex solution with a unit topological charge \cite{Kondo25} is given in Fig~\ref{F_vortex}.
%%%%%%%%%%%%%%%%%%%%%%%%%%%%%%%%%%%%%%%%%%%%%%%%%%%%%%%%%%%
%%%%%%%%%%%%%%%%%%%%%%%%%%%%%%%%%%%%%%%%%%%%%%%%%%%%%%%%%
%\end{comment}

\noindent
\textbf{Proposition} [area law of the Wilson loop average\cite{Kondo25}]
In the %\textcolor{red}
{dilute gas approximation}, the Wilson loop average in the $\vartheta$ vacuum %(\ref{Wilson_loop_A}) 
obeys the %\textcolor{red}
{area law} with the area $A(C)$ in the semi-classical regime: 
\begin{align}
\langle \vartheta | W_C[\mathscr{A}] | \vartheta \rangle = e^{- \sigma A(C)} , \quad
%= \exp \left\{ - 2K e^{-S_1/\hbar}
%\left[ \cos \vartheta c_2- \cos \left(\vartheta c_2+ 2\pi J c_1 \right)
%\right]A(C) \right\} .
%%\nonumber\\
\sigma := 2K e^{-S_1/\hbar} \left[ \cos (\vartheta c_2) - \cos
\left(\vartheta c_2 + 2\pi J c_1\right) \right] ,
\label{wler}
\end{align}
where $c_1$ and $c_2$ are the first and second Chern numbers corresponding to the vortex number and the Yang-Mills instanton number respectively. 
Here %$\vartheta$ denote the topological angle, 
$S_1$ is the 1-vortex action, $J=\frac12, 1, \frac32, \cdots$ is the index for the representation, and $K$ is the fugacity of the dilute  (instanton) gas. 
%We note here that the volume dependence disappears by taking the ratio $I_2^\vartheta/I_1^\vartheta$.
%Thus, the Wilson loop expectation value in a vacuum with the topological angle $\vartheta$ shows the area law.
%Taking the rectangular Wilson loop leads to the static quark potential:
%\begin{equation}
%V(R) = \sigma R , \quad
%\sigma = 2K e^{-S_1/\hbar} \left[ \cos (\vartheta c_2) - \cos \left(\vartheta c_2 + 2\pi J c_1\right) \right] .
%\label{pot}
%\end{equation}
%\\
When $Jc_1$ is an integer, the vacuum is periodic with respect to $\vartheta c_2$ with period $2\pi$, so the potential is zero. 
%This means that the integral charges are screened by the formation of neutral bound states.
When $Jc_1$ is not an integer, % (when $q$ is not an integer multiple of the elementary charge $g$), 
the static $q\bar q$ potential $V(R)$ has a linear potential $V(R)=\sigma R$ with the string tension $\sigma$.% as the proportionality coefficient.

\section{Conclusions and discussions}
%\textcolor{green}{\Large\bf $\S$~~Conclusions and discussions
%}
%\setcounter{equation}{0}
%\normalsize 
%\small

%\noindent
%Conclusion:
%\\
%\noindent
%$\bullet$ 
We considered the space and time {symmetric instantons} as solutions of the {self-dual Yang-Mills equation with conformal symmetry} in the $SU(2)$ Yang-Mills theory in the four-dim.  Euclidean space $\mathbb{R}^4$. 
%\noindent
%$\bullet$ 
%The instanton with \textcolor{red}{time translation symmetry} (time independence) is the well-known Prasad-Sommerfield (PS) magnetic monopole with a finite energy.  However, the PS solution gives an infinite four-dim.  action.  Therefore it gives no contribution in  the path integral and is not responsible for quark confinement. %plays no role in the quantum Yang-Mills theory. 
%\noindent
%$\bullet$ 
%On the other hand, 
In contrast to time translation symmetry, 
instantons with {spatial rotation symmetries} give a finite four-dim.  action and hence can contribute to quark confinement. %in the quantum Yang-Mills theory. 
%\\
For the {spatial symmetry} $SO(2) \simeq U(1) \simeq S^1$, the instanton is reduced to a {hyperbolic magnetic monopole} (of Atiyah) living in the three-dim.  {hyperbolic space} $\mathbb{H}^3$.
%\\
For the {spatial symmetry} $SO(3) \simeq SU(2)$, the instanton is reduced to a {hyperbolic vortex} (of Witten-Manton) living in the two-dim.  {hyperbolic space} $\mathbb{H}^2$. 

%\noindent
%$\bullet$ 
By requiring the spatial symmetry $SO(2)$ or $SO(3)$ for instantons,  the four-dim.  Euclidean space $\mathbb{R}^4$ %in which instantons live 
is inevitably mapped %transformed
 to the curved space $\mathbb{H}^3 \times S^1$ or $\mathbb{H}^2 \times S^2$ with negative constant curvature by maintaining the {conformal equivalence} through {dimensional reduction}. 
%As a result, the spacetime was partially compactified.
%The spatial symmetry $SO(2)$ or $SO(3)$ acts on the compact space $S^1$ or $S^2$.
%\noindent
%$\bullet$
%On these curved spacetimes, we considered instantons with spacetime symmetry acting on the compact spaces $S^1$ and $S^2$, that is, $SO(2)$-symmetric instantons and $SO(3)$-symmetric instantons.
%As a result, dimensional reduction occurred, and we reduced them to Atiyah's three-dimensional hyperbolic magnetic monopole and Witten-Manton's two-dimensional hyperbolic vortex, respectively.
%\noindent
%$\bullet$
%Three-dimensional
{Hyperbolic magnetic monopoles} on $\mathbb{H}^3$ and  {hyperbolic vortices} on $\mathbb{H}^2$ can be connected through {conformal equivalence} with the explicit relationship between the magnetic monopole field and the vortex field has been obtained, which allows magnetic monopoles and vortices can be treated in a unified manner.

%\noindent
%$\bullet$
Both $\mathbb{H}^3$ and $\mathbb{H}^2$ are curved spaces $AdS_3$ and $AdS_2$ with constant negative curvatures. 
The \textbf{hyperbolic monopole in $\mathbb{H}^3$ is completely determined by its holographic image on the conformal boundary two-sphere $S_\infty^2$}. 
(This is different from Euclidean monopoles.)
This fact enables us to reduce the non-Abelian Wilson loop operator to the Abelian Wilson loop defined by the Abelian gauge field of the vortex: 
{Abelian dominance} and {magnetic monopole dominance}.

%\noindent
%$\bullet$
Using the hyperbolic magnetic monopole and hyperbolic vortex obtained in this way, quark confinement was shown to be realized in the sense of {Wilson area law} within the {dilute gas approximation}.
This is a semi-classical quark confinement mechanism originating from the unified hyperbolic magnetic monopole and hyperbolic vortex, supporting the {dual superconductor picture}.

%\noindent
%[
%$\bullet$
%Furthermore, by considering a symmetric instanton with a singularity (of Forgacs-Horvath-Palla(1981)) in a compact subspace of spacetime, a symmetric instanton with a \textcolor{red}{non-integral topological charge} can be obtained, and then by dimensional reduction, a hyperbolic magnetic monopole and a hyperbolic vortex with a non-integral topological charge have been obtained.
%]

\begin{comment}
%\newpage
\noindent
Discussion:

\noindent
$\bullet$
Why does the space-time obtained by dimensional reduction have negative curvature? Is there no case where it has positive curvature?
cf: The 4-dimensional standard model can be obtained by dimensional reduction of 6-dimensional Yang-Mills theory to 4!
[Manton(1981)]

\noindent
$\bullet$
How does the gauge group change due to dimensional reduction?

\noindent
$\bullet$
How can it be extended to a large gauge group $SU(N)$?

\noindent
$\bullet$
What happens when a matter field is introduced? For example, can QCD be analyzed in the same way?

\noindent
$\bullet$
How do we incorporate quantum effects that do not maintain conformal invariance?

\end{comment}

%\appendix
%\section{}

%Use the \verb|\appendix| command if you need an appendix(es). The \verb|\section| command should follow even though there is no title for the appendix (see above in the source of this file).

\section*{Acknowledgements}
This work was supported by Grant-in-Aid for Scientific Research, JSPS KAKENHI Grant
Number (C) No.23K03406.

\end{document}